\documentclass[11pt,fleqn]{article}
\usepackage{amsmath,amssymb} 
\usepackage{graphicx}

\newcommand{\eps}{\varepsilon}
\newcommand{\R}{\mathbb{R}}
\newcommand{\dee}{\mathrm{d}}
\newcommand{\imag}{\mathrm{i}}
\newcommand{\rem}[1]{}
\newcommand{\noalph}{\vartheta}

\begin{document}
\title{Vanishing Twist in the Hamiltonian Hopf Bifurcation}
\author{H.~R.~Dullin$^{1, 2}$, A.~V.~Ivanov$^{1}$
\thanks{This research is funded by the EPSRC under contract GR/R44911/01. 
Partial support by the European  Research Training Network
{\it Mechanics and Symmetry in Europe\/} (MASIE), HPRN-CT-2000-00113,
is also gratefully acknowledged. AVI was also supported in part by INTAS grant 00-221, RFBI grant
01-01-00335 and RFME grant E00-1-120.}
\\
\\ $^{1}$ Department of Mathematical Sciences, 
\\ Loughborough University,  LE11 3TU, UK
\\ $^{2}$ Fachbereich 1, Physik, Universit\"at Bremen
\\ 28334 Bremen, Germany
\\ {\small H.R.Dullin@lboro.ac.uk},  {\small A.V.Ivanov@lboro.ac.uk}
}
\maketitle

\begin{abstract}

The Hamiltonian Hopf bifurcation has an integrable normal form 
that describes the passage of the eigenvalues of an equilibrium 
through the $1: -1$ resonance. At the bifurcation the pure imaginary 
eigenvalues of the elliptic equilibrium turn into a complex quadruplet
of eigenvalues and the equilibrium becomes a linearly unstable
focus-focus point.
We explicitly calculate the frequency map of the integrable normal form, 
in particular we obtain the rotation number as a function
on the image of the energy-momentum map in the case where the 
fibres are compact. We prove that the isoenergetic 
non-degeneracy condition of the KAM theorem is violated on a curve
passing through the focus-focus point in the image of the energy-momentum 
map. This is equivalent to the vanishing of twist in a Poincar\'e map 
for each energy near that of the focus-focus point. In addition we show that
in a family of periodic orbits (the non-linear normal modes) the twist 
also vanishes.
These results imply the existence of all the unusual dynamical 
phenomena associated to non-twist maps near the Hamiltonian
Hopf bifurcation.


\vspace*{1ex}

\end{abstract}


\section{Introduction}

Understanding the dynamics of a Hamiltonian system near equilibrium points
is of fundamental importance. In the elliptic case the eigenvalues of the 
linearization are pure imaginary, $\lambda_j = \pm i\omega_j$, $j=1,\dots,n$,
where $n$ is the number of degrees of freedom, which will be 2 in the following. 
By the Lyapunov Centre theorem
the $n$ normal modes of the linear approximation persist in the non-linear 
system when the eigenvalues are non-resonant.
The resonant cases were much more recently treated by
\cite{Weinstein73,Moser76,Duistermaat84,MRS88}. 
In some sense the most exceptional resonance
is the so called $1:-1$ resonance, in which the quadratic part $H_2$ of the 
Hamiltonian $H$ has degenerate eigenvalues and is not definite, 
$2 H_2 = \omega (p_x^2 + x^2)  - \omega (p_y^2 + y^2)$.
The unfolding of the normal form gives a family of Hamiltonian systems
with an equilibrium point that looses stability by passing through the 
$1:-1$ resonance. This is called the Hamiltonian Hopf bifurcation, 
and was studied in detail by \cite{Sokolskii77,vanderMeer85}, 
also see \cite{AKN,Duistermaat98}. 
The corresponding family of Hamiltonians is Liouville integrable.
In this paper we show that near this bifurcation the twist vanishes.
This means that the rotation number (i.e.\ the ratio of frequencies of 
the Hamiltonian flow on invariant tori)
of the one parameter family of invariant tori with fixed energy 
has a critical point. 

The dynamical consequences of vanishing twist are well known. 
They were first described by \cite{HowHoh84}, and later studied
by \cite{HowHum95, CGM96, Simo98}. 
In \cite{DMS98b} we have shown that the vanishing of twist
in one parameter families of maps occurs near the $1:3$ resonance,
also see \cite{Moeckel90}.
In \cite{DM02} we have shown that also in 4 dimensional symplectic maps
the vanishing of twist appears near resonance.
More recently in \cite{DI03a} we have shown that the twist also vanishes
near the saddle-centre bifurcation, in which one multiplier passes through 
zero. In this paper we show that the principle that the twist vanishes
near resonance also applies in the Hamiltonian 
Hopf bifurcation. For flows the condition of vanishing twist
is one of the conditions for the standard form of the KAM theorem to 
hold. In this setting it is usually called the isoenergetic 
non-degeneracy condition. There exist KAM theorems
with weaker conditions \cite{DL98,Ruessmann87}, so that 
vanishing twist does not necessarily imply that the torus will be destroyed. 
It does mean, however, that a resonant twistless torus will create
all the unexpected dynamics described by the twistless standard map.

In the following two section we present well known material about the
Hamiltonian Hopf bifurcation, in order to introduce the Hamiltonian
and its Energy-Momentum map and to fix our notation. 
Then our own contribution starts with the 
derivation of  the actions and the rotation number. The rotation number
and its derivative are analysed near critical values of the energy-momentum
map, namely near the isolated focus-focus point in the compact case 
and on the family of relative equilibria. The details of the expansion of the
elliptic integrals are given in the appendix.

\goodbreak

\section{Hopf Normal Form}

Consider coordinates $q = (q_{1}, q_{2})$ and conjugate momenta $p = (p_{1}, p_{2})$
so that the symplectic form on $\R^4$ is
$\Omega = \dee p_1 \wedge \dee q_1 + \dee p_2 \wedge \dee q_2$.
The normal form for the Hamiltonian Hopf bifurcation is
\begin{equation} \label{Ham1}
H(q_{1},q_{2},p_{1},p_{2}) = \beta \Gamma_{1} + \Gamma_{2} + \delta (\gamma \Gamma_{1} + \Gamma_{3}) + C \Gamma_{1}^{2} + 2 B  \Gamma_{1}\Gamma_{3} + 2 D \Gamma_{3}^{2} + O_{3}(\Gamma_{1}, \Gamma_{2}, \Gamma_{3}),
\end{equation}
where $\Gamma_{1} = p_{2} q_{1} - p_{1} q_{2}$, $\Gamma_{2} = \frac{1}{2}(p_{1}^{2} + p_{2}^{2})$, $\Gamma_{3} = \frac{1}{2}(q_{1}^{2} + q_{2}^{2})$, and $\delta$ is a bifurcation parameter, $\beta, \gamma, B, C, D$ are real constants such that $\beta \ne 0$ and $D\ne 0$. The expression $O_{3}$ denotes terms of order no less than 3 with respect to $\Gamma_{i}, i = 1,2,3$. For simplicity we will use the notation $\omega = \beta + \delta \gamma$. 
The system has an equilibrium point at the origin $p_i = q_i = 0$ with
eigenvalues $\sqrt{-\delta} \pm i\omega$. 
For ease of notation we write $\alpha = \sqrt{-\delta}$ when $\delta < 0$, 
so that the eigenvalues of the equilibrium are $\pm \alpha \pm \imag \omega$.
The dependence of $\omega$ on 
$\delta$ is not essential for our purposes, because $\beta \not = 0$. 
By a symplectic scaling with multiplier the parameter $\omega$ can be
scaled to 1, and $D$ can be scaled to $\pm 1$ at the same time. We find
it useful, however, to keep unscaled variables and parameters until the very last section.

The Hamiltonian system (\ref{Ham1}) is Liouville integrable. A second independent 
constant of motion is $\Gamma_1$. It generates the $S^1$ symmetry
\begin{equation}
\Phi : S^{1}\times \mathbb{R}^4 \to \mathbb{R}^{4},\ \ \  \Phi(\noalph, q, p) = (S_{\noalph}q, S_{\noalph}p), \quad
S_{\noalph} = \begin{pmatrix}  \cos \noalph & \sin \noalph \\ -\sin \noalph & \cos \noalph \end{pmatrix}
\end{equation}
The corresponding momentum map $J : \R^4 \to \R$ is given by 
$J(q,p) = p_{2} q_{1} - p_{1} q_{2}$, which is $\Gamma_1$. 
Since $J$ generates the periodic flow $\Phi$ with period $2\pi$
it is an action of the integrable system. We denote the (constant) value of $J$ by $j$.
To perform the reduction with respect to this symmetry we use invariant
theory, see e.g.~\cite{CushmanBates97}. Singular reduction occurs in this
example because the action $\Phi$ is not free: the equilibrium 
(= the origin) is a fixed point of this action.
The algebra of polynomials in $\R^4$ that are
invariant under $\Phi$ is generated by $\Gamma_1$, $\Gamma_2$, 
$\Gamma_3$ and $\Gamma_{4} = p_{1}q_{1} + p_{2}q_{2}$. 
This means that any polynomial of $q_1$, $q_2$, $p_1$, $p_2$ that is 
invariant under $\Phi$ can be written as a polynomial of $\Gamma_i$, $i=1,\dots,4$.
The generators satisfy the relations
\begin{equation} \label{Genrel}
G(\Gamma) = \Gamma_{1}^{2}/2 + \Gamma_{4}^{2}/2 - 2\Gamma_{2}\Gamma_{3} = 0, \ \  \Gamma_{2} \ge 0, \ \  \Gamma_{3} \ge 0.
\end{equation}
The reduced phase space $P_j = J^{-1}(j)/S^1$ is defined by (\ref{Genrel}) with $\Gamma_1 = j$
as a semialgebraic variety in $\R^3$ with coordinates $(\Gamma_2, \Gamma_3, \Gamma_4)$. 
If $j\neq 0$ the reduced phase space $P_j$ is one sheet of a two-sheeted hyperboloid given
by (3), so it is a smooth manifold. But for $j=0$ it is half of an elliptic cone and hence is not smooth because of the singular point of the cone at the origin 
$(\Gamma_{2}, \Gamma_{3}, \Gamma_{4}) = (0, 0, 0)$. 

\begin{figure}
\centerline{\includegraphics[width=8cm]{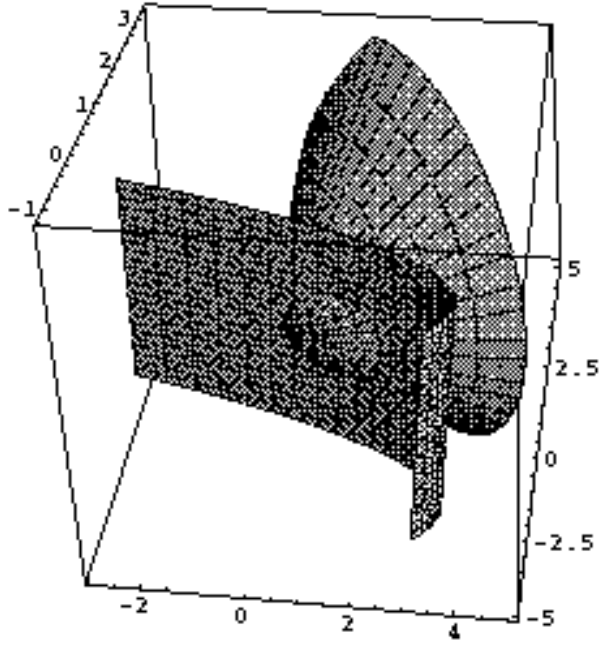}}
\caption{A typical intersection $M_{h,j}$ of the surfaces 
$P_{j}$ and $H_j^{-1}(h)$. \label{figred}}
\end{figure}

The reduced Hamiltonian is
\begin{equation} \label{HamRed}
H_j(\Gamma_2,\Gamma_3,\Gamma_4) = \omega j + \Gamma_{2} + \delta \Gamma_{3} + 
C j^2 + 2 B  j \Gamma_{3} + 2 D \Gamma_{3}^{2}\,.
\end{equation}
The surface $H_j^{-1}(h)$ is a parabolic cylinder in $(\Gamma_2,\Gamma_3,\Gamma_4)$
that is independent of $\Gamma_4$. The integral curves of the reduced system
are given by the intersection of the surface $P_j$ with the surface $H_j^{-1}(h)$, 
as illustrated in Fig.~\ref{figred}. We denote the intersection by $M_{h,j}$.

The Poisson bracket $\{\cdot , \cdot\}$ associated with the standard symplectic structure $\Omega$ on $\mathbb{R}^{4}$ defines a Poisson structure on the algebra of invariant polynomials with brackets
\begin{equation}
\{\Gamma_{1}, \Gamma_{2}\} = 
\{\Gamma_{1}, \Gamma_{3}\} = \{\Gamma_{1}, \Gamma_{4}\} = 0, 
\end{equation}
\begin{equation}
\{\Gamma_{2}, \Gamma_{3}\} =\Gamma_{4}, \quad
\{\Gamma_{3}, \Gamma_{4}\} = -2\Gamma_{3}, \quad
\{\Gamma_{4}, \Gamma_{2}\} = -2\Gamma_{2}\,. 
\end{equation} 
The momentum map of the $S^1$ action used for the reduction induces the Casimir
$\Gamma_1$ in this Poisson bracket. In addition also the relation between 
the generators (\ref{Genrel}) given by $G$ is a Casimir. Accordingly the nonzero brackets
give a Poisson structure on $\R^3$ with coordinates $(\Gamma_2, \Gamma_3, \Gamma_4)$,
that has the reduced spaces $P_j$ as its symplectic leaves.
It  can be written as
\begin{equation}
\{\Gamma_{2}, \Gamma_{3}\} = \partial G / \partial \Gamma_4,  \quad
\{\Gamma_{3}, \Gamma_{4}\} = \partial G / \partial \Gamma_2, \quad
\{\Gamma_{4}, \Gamma_{2}\} = \partial G / \partial \Gamma_3\,.
\end{equation} 
The reduced equations of motion are
\begin{equation}
\begin{aligned}
\dot \Gamma_2 &= \{\Gamma_2, H_j \} = \Gamma_4 \frac{\partial H_j }{ \partial \Gamma_3}, \\
\dot \Gamma_3 &= \{\Gamma_3, H_j \} = -\Gamma_4, \\
\dot \Gamma_4 &= \{\Gamma_4, H_j \} = -2\Gamma_2 + 
       2 \Gamma_3 \frac{ \partial H_j }{ \partial \Gamma_3 }\,.
\end{aligned}
\end{equation}
The integral curves of this flow are given by $M_{h,j}$, the intersection of 
$P_j$ and $H_j^{-1}(h)$. 
In general the intersection of these two manifolds is either empty or diffeomorphic to 
a circle. 
The preimage of any point in reduced phase space 
is the set of points in original phase space that are
mapped to this point by the momentum map $J$.
If one point in the preimage is known the others can be obtained by letting 
the flow of $J$ 
(i.e.\ the map $\Phi$) act on this point to get the complete fibre.
This gives a circle unless starting in the origin.
Therefore the preimage of a circle $M_{h,j}$ 
is a two dimensional torus $\mathbb{T}^{2}$ in the phase space   of the original system. 

Exceptions occur for equilibrium points of the reduced system.
They occur either when the surface $H_j^{-1}(h)$ is tangent to $P_{j}$ 
or when $j=0$ and $H_0^{-1}(h)$ contains the singular point at the origin 
(which implies $h=0$). 
The preimage of the singular point is not a circle, because $\Gamma_2 = \Gamma_3 = 0$
implies $q_1 = q_2 = p_1 = p_2 = 0$ and this is a fixed point of the flow $\Phi$.
This is the equilibrium point in the full system that undergoes the Hopf bifurcation.
All other equilibrium points of the recuded system are reconstructed to periodic
orbits of the full system; they are relative equilibria of $J$.
The $S^1$ action generated by $J$ is not free. The origin is a fixed point, and 
this is the reason why singular reduction is needed in this example.

\section{Energy-Momentum Map}

Using the reduced system we can find the critical values of the energy
momentum map 
\begin{equation}
  F : \R^4 \to \R^2, \qquad (p,q) \mapsto F(p,q) = ( H(p,q), J(p,q) )\,.
\end{equation}
The values of the energy-momentum map are denoted by $(h,j)$. For every
regular value of $F$ the preimage in phase space is a two dimensional torus.
The critical values are determined from equilibrium points of the reduced system
because their preimages are not $\mathbb{T}^2$.
Since we are interested in a neighbourhood of the origin in phase space
for small $\delta$ we will only consider a small neighbourhood of the origin 
in the image of the momentum map.

Consider the reduced equilibrium points caused by the singularity in the 
reduced space first. This singularity occurs for $j=0$.
The singular point $(\Gamma_{2},\Gamma_{3}, \Gamma_{4}) = (0, 0, 0)$ has energy $h=0$.
The equilibrium at the origin in phase space is therefore mapped to the 
origin in the image of the momentum map.

When $\delta > 0$ the intersection $M_{0,0}$ restricted to a neighborhood of the
origin in reduced phase space consists only of the origin. 
It reconstructs to an elliptic equilibrium.
However, if $\delta < 0$ then $M_{0,0}$ is a non-smooth circle with a corner, 
if it is compact. 
The preimage of $M_{0,0}$ is diffeomorphic to a pinched torus in this case.

Consider next the equilibrium points caused by a tangency of 
$P_j$ and $H_j^{-1}(h)$.
At these critical values of $H_j$ the gradient of $H_j$ and the gradient
of $G$ are parallel. Since $\partial G/\partial \Gamma_4 = \Gamma_4$
the tangency may occur only on the hyperplane $\Gamma_{4} = 0$. 
The intersections of $P_j$ and $H_j^{-1}(h)$ with this hyperplane are 
one branch of a hyperbola and a parabola, respectively. 
They are described by the equations
\begin{align}  \label{hyperbola}
j^2 & = 4 \Gamma_{2}\Gamma_{3}, \\
h &= \omega j + Cj^2 + \Gamma_2 - \frac{(\delta + 2Bj)^2}{8D} + 
     2D\left(\Gamma_3 + \frac{\delta + 2Bj}{4D}\right)^2\,. \label{parabola}
\end{align} 
At the extremal values of $h$ the two curves are tangent. 
Eliminating $\Gamma_{2}$ in (\ref{hyperbola}) using (\ref{parabola}) gives a polynomial of
degree 3 in $\Gamma_{3}$ depending on $j$ and $h$
\begin{equation} \label{eqn:Q3}
Q_3(\Gamma_3) := -8D\Gamma_{3}^{3} - 4 (\delta + 2 B j)\Gamma_{3}^{2} + 
        4 (h - \omega  j - C j^{2} )\Gamma_{3} - j^{2}  = 0\,.
\end{equation}
This polynomial $Q_3(\Gamma_3)$ gives the value of $\Gamma_4^2$ obtained
from $G=0$ and expressed in terms of $\Gamma_3$.
The tangency between the hyperbola (\ref{hyperbola}) and the 
parabola (\ref{parabola}) occurs when $Q_3$ has a double root.
We will first discuss all values of $(h,j)$ for which 
a tangency occurs, irrespective of them satisfying the 
constraints $\Gamma_2\ge 0$ and $\Gamma_3 \ge 0$.
In a second step the critical values of the energy 
momentum will be found by consideration of these constraints.

\begin{figure}
\centerline{\includegraphics[width=10cm, height=6cm]{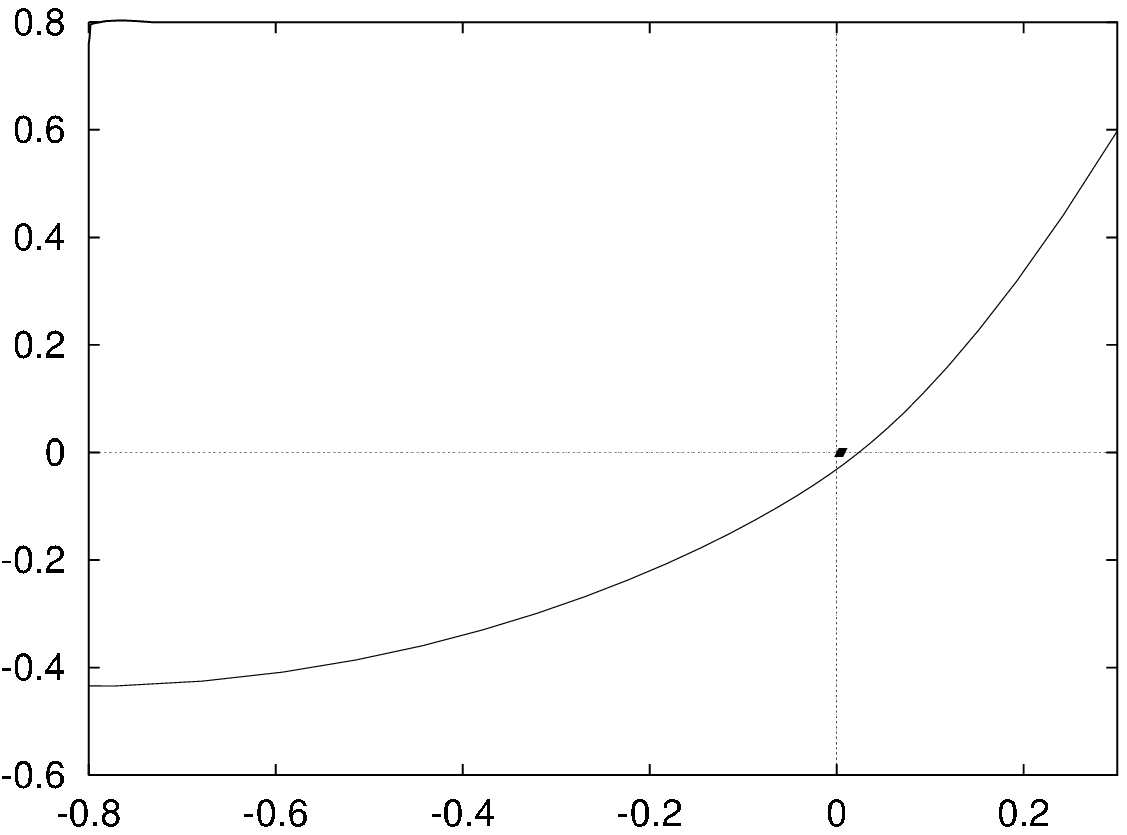}}
\centerline{\includegraphics[width=10cm, height=6cm]{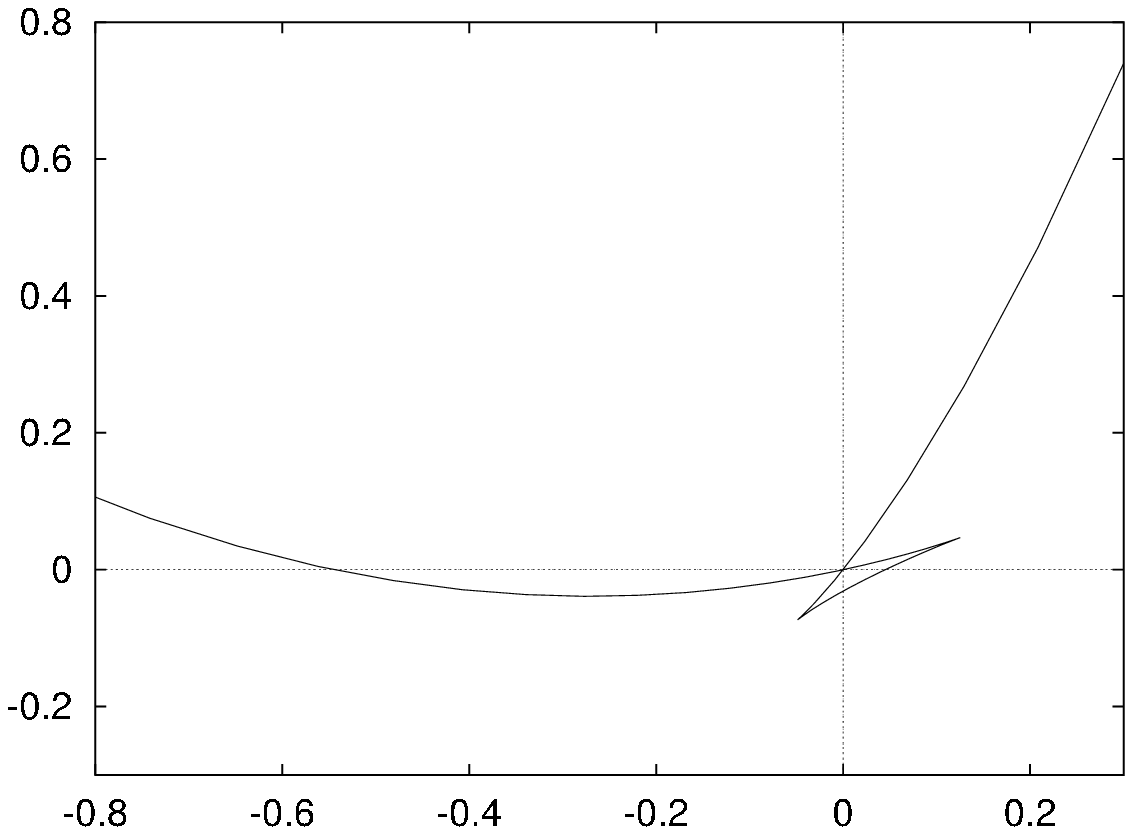}}
\caption{
The discriminant surface of the polynomial $Q_3$. 
The critical values of the energy-momentum map 
are contained in the discriminant surface.
The small triangular part for $\delta > 0$ is not in the image of
the energy-momentum map for $D< 0 $.
The figures correspond to the cases of $\delta = -0.15$ and $\delta = 0.15$ respectively. 
($B = C = -D = \omega = 1.0$)} \label{fig:bifdiag}
\end{figure}

To parametrize all tangencies we make the ansatz 
$Q_3(z) = -8D(z-d/2)^2(z+s^2/D/2)$ with parameters $d$ and $s$ 
parametrising the double and single root of $Q_3$, respectively.
This leads to the parametrisation of the tangencies by $s \in \R$
\begin{equation} \label{hjofs}
\begin{aligned}
j_\text{crit}(s) & = sd(s), \qquad
d(s) = \frac{\delta - s^2}{2(Bs - D)}, \quad \\
h_\text{crit}(s)  & = \omega j_\text{crit}(s)  + Cj_\text{crit}(s)^2 - 2d(s)^2D + 2sj_\text{crit}(s)  \,. 
\end{aligned}
\end{equation}
The root $\Gamma_3 = -s^2/D/2$ always has the opposite sign than $D$. 
The curve $(h_\text{crit} (s),j_\text{crit} (s))$ has singular points when $s$ has one of the 
singular values satisfying $2Bs^3 - 3s^2 D + \delta D = 0$.
The number of singular points changes when the discriminant 
$108 \delta D^2 ( D^2 - \delta B^2)$ vanishes. 
For small $|\delta|$ the only change occurs at $\delta = 0$, see Fig.~\ref{fig:bifdiag}, 
for two slices of the ``swallowtail''.

For $\delta > 0$ the curve has two singular points near the origin 
for some $s \in (-\sqrt{\delta}, \sqrt{\delta})$.
The two singular points are located at
\begin{equation}
\begin{aligned}
h^* &= \pm \frac{\omega}{\sqrt{27}D} \delta^{3/2}  + \frac{2B\omega-3D}{18D^2}\delta^2 
     \pm O(\delta^{5/2}) \\
j^* &= \pm \frac{1}{\sqrt{27}D} \delta^{3/2}  + \frac{B}{9D^2}\delta^2 \pm O(\delta^{5/2})
\end{aligned}
\end{equation}
for small $\delta \ge 0$.
The curve of critical values has a self-intersection at $s^2 = \delta$.
The intersection point at the origin marks the elliptic equilibrium with eigenvalues
$i (\omega \pm \sqrt{\delta})$. The slopes of the intersecting curves are given by 
the imaginary parts of the eigenvalues. 
  
For $\delta < 0$ the equilibrium point at the origin is
unstable with eigenvalues $\sqrt{-\delta} + i \omega$. The curve 
$(h_\text{crit} (s),j_\text{crit} (s))$ does 
not intersect the origin, instead the origin is now an isolated critical point. 
The curve is above the origin for $D > 0$ and below for $D < 0$, e.g.\ the 
point with $s=0$ is at $(h,j) = (-\delta^2/8/D, 0)$.

Only tangencies that occur with the part of the hyperbola in the positive 
quadrant give critical values of the energy-momentum map since 
$\Gamma_2$ and $\Gamma_3$ are both non-negative.
The double roots $d(s)$ near $s=0$ occurs at
\begin{equation}
 d(s) = -\frac{\delta}{2D} - O(s)\,.
\end{equation}
Near the intersection at the origin around $s = \pm \delta$ the double root is
\begin{equation}
 d(s) =  \frac{\sqrt{\delta}}{D}(\pm s - \sqrt{\delta}) + O( (\pm s - \sqrt{\delta})^2 )
\end{equation}
If $D>0$ this implies that the smooth curve for $\delta < 0$ is in the (boundary of the)
image of the energy-momentum map, 
while the part of the curve with $s \in (-\sqrt{\delta}, \sqrt{\delta})$ 
is {\em not} in the image for $\delta > 0$.
Conversely for $D<0$ the smooth curve for $\delta < 0$ is not in the image, 
while for $\delta > 0$ only the triangular part with 
$s \in [-\sqrt{\delta}, \sqrt{\delta}]$ is in the image.
Therefore the union of the bifurcation diagram (i.e.\ the set of critical values of the
energy-momentum map) for $+D$ and $-D$ gives
the discrimiant of the polynomial $Q_3$. 

The type of the preimage of the critical values is determined by the character of 
the intersection $M_{h,j}$. The positive half of the hyperboloid $G$ projects onto the area 
above the hyperbola given by $j^2 \le 4\Gamma_2\Gamma_3$. 
If the parabola (\ref{parabola}) touches the
boundary of the area from the outside, the preimage in the full phase space is
a circle, hence a stable periodic orbit. If the parabola touches from
the inside, the preimage is a circle with a separatrix, hence an unstable periodic orbit.
This can only occur when $D<0$, because then the parabola is open upwards.
It only occurs when $\delta > 0$ for $s$ between the two singular values enclosing zero.
In all other cases the parabola touches from the outside.
The complete bifurcation scenario in the two cases therefore is as follows, 
see Figure~\ref{fig:bifdiag} for illustration.

{\em The case $D > 0$:} 
For $\delta < 0$ there is an isolated focus-focus point 
at the origin and a smooth curve of elliptic periodic orbits nearby.
For $\delta > 0$ there is an elliptic equilibirum point 
and there are two families of elliptic periodic orbits (non-linear normal modes) 
emanating from the equilibrium.

{\em The case $D < 0$:} 
For $\delta < 0$ there is nothing but an isolated focus-focus point.
For $\delta > 0$ there is an elliptic equilibirum point 
and there are two families of elliptic periodic orbits (non-linear normal modes) 
emanating from the equilibrium. Both families terminate in a cusp formed with the same family 
of hyperbolic periodic orbits. The set of critical values therefore forms a triangle with two 
cuspoidal corners and one regular corner at the origin.

\section{Actions}

From now on we shall assume the parameter $D$ to be a positive number. In this case each constant energy level is compact and by the Liouville-Arnold theorem 
it is possible to define action-angle coordinates near regular points of $F$.
To construct the second action we need to integrate a 1-form $\Theta$ over $M_{J,h}$, where the differential ${\rm d}\Theta$ coincide with a symplectic structure induced by the quotient map $\pi_{h,j}: \R^{4} \to F^{-1}(h,j)/S^{1}$ from the original phase space, i.e. ${\rm d}\Theta = (\pi_{h,j}^{-1})^{\*}\Omega\vert_{TM_{h,j}}$.
To find the form $\Theta$ we chose $\Gamma_3$ as one variable and
find its conjugate variable by solving the equation
\begin{equation} \label{PBactPDE}
\{f(\Gamma_{2}, \Gamma_{3}, \Gamma_{4}), \Gamma_{3}\} = 1.
\end{equation}
A solution of (\ref{PBactPDE}) is the function 
\begin{equation}
f(\Gamma_{2}, \Gamma_{3}, \Gamma_{4}) = \frac{\Gamma_{4}}{2\Gamma_{3}}.
\end{equation}
Then the canonical one form is
$\Theta = \frac{\Gamma_{4}}{2\Gamma_{3}}{\rm d} \Gamma_{3}$ and we obtain
\begin{equation} \label{ActGamma}
J_2(h,j) 
= \frac{1}{2\pi}\oint\limits_{M_{h,j}} \frac{\Gamma_{4}{\rm d}\Gamma_{3}}{2\Gamma_{3}}
= \frac{1}{2\pi}\oint\limits_{M_{h,j}} \frac{\sqrt{4\Gamma_2\Gamma_3-j^2}}{2\Gamma_3} \dee \Gamma_3
\end{equation}
for the second action. Here $\Gamma_4$ is considered as a function of $\Gamma_3$
by first expressing $\Gamma_4$ in terms of $\Gamma_2$ and $\Gamma_3$ on the
reduced phase space $P_j$ and
then by expressing $\Gamma_2$ in terms of $\Gamma_3$ using $H_j = h$.
As a result the polynomial $Q_3$ is found as already given by (\ref{eqn:Q3}). 
The action integral hence is defined on the elliptic curve
\begin{equation} \label{Ecurve}
   {\cal E}  = \{ (w,z) : w^2 = Q_3(z) \}\,.
\end{equation}
Recall that $w = \Gamma_4$ on $P_j$.
Now the action integral can be written as
\begin{equation}
  J_2 = \frac{1}{4\pi } \oint \frac{w}{z} \dee z \,.
\end{equation}
It is an integral of the third kind with a pole at $z=0$ and residue $\pm \imag j/4\pi$.

The formula for the action can also be obtained in a classical way, 
using polar coordinates as in \cite{Sokolskii77}.
\rem{ 
The are $\Omega = \dee r \wedge \dee P + \dee \phi \wedge \dee J$ and
\begin{equation}
q_{1} = r \cos \varphi, \quad p_{1} = P \cos \varphi - J \frac{\sin \varphi}{r}, 
\end{equation}
\begin{equation}
q_{2} = r \sin \varphi, \quad p_{2} = P \sin \varphi + J \frac{\cos \varphi}{r}.
\end{equation}
After this transformation the Hamiltonian of the system takes the form
\begin{equation}
H = \frac{1}{2}(P^{2} + J^{2}/r^{2}) + 
\omega J + \frac{1}{2} (\delta + 2 B J)r^{2} + C J^{2} + \frac{1}{2} D r^{4}\,.
\end{equation}
The Hamilonian does not depend on the coordinate $\varphi$, so the conjugate momentum $J$ is a constant of motion and in fact it is one of the action variables $J_{1}$ of the system. To calculate another action variable $J_2$ express the momentum $P$ from the equation of the constant energy level and substitute it into the formula
\begin{equation}
J_2 = \frac{1}{2\pi}\oint P {\rm d}r.
\end{equation}
Thus we get
\begin{equation}
J_2 = \frac{1}{2\pi}\oint \frac{\sqrt{2(h - \omega J - C J^{2}) r^{2} - (\delta + 2 B J) r^{4} - D r^{6} - J^{2}}}{r}{\rm d}r.
\end{equation}
This is equivalent to (\ref{ActGamma}) upon transforming $r^2 = 2 \Gamma_3$.
} 
A slightly different coordinate transformation illucidates the connection between
the two approaches. The new symplectic structure is
$\Omega = \dee P_g  \wedge \dee g +  \dee J \wedge \dee \phi$ and 
``symplectic polar coordinates'' valid for $q_1^2 + q_2^2 > 0$ are introduced by 
\begin{equation}
q_{1} = \sqrt{2g} \cos \varphi, \quad 
p_{1} = P_g \sqrt{2g} \cos \varphi - J \frac{\sin \varphi}{\sqrt{2g}}, 
\end{equation}
\begin{equation}
q_{2} = \sqrt{2g} \sin \varphi, \quad 
p_{2} = P_g \sqrt{2g} \sin \varphi + J \frac{\cos \varphi}{\sqrt{2g}}.
\end{equation}
The invariant polynomials are related to these coordinates by 
\begin{equation}
\Gamma_1 = J, \quad \Gamma_2 = g P_g^2 + \frac{J^2}{4g}, \quad
\Gamma_3 = g, \quad \Gamma_4 = 2 g P_g\,.
\end{equation}
In these variables the Hamiltonian takes the form
\begin{equation} \label{eq:Hamg}
H = gP_g^{2} + \frac{J^{2}}{4g} + 
\omega J +  (\delta + 2 B J)g + C J^{2} + 2 D g^2\,,
\end{equation}
and the equations of motion are
\begin{equation} \label{phidot}
\begin{aligned}
\dot \varphi &= \frac{J}{2g} + \omega + 2Bg +  2CJ \,, \\
\dot      g &= 2 g P_g \,, \\
\dot P_g &= P_g^2 - \frac{J^2}{4g^2} + (\delta + 2 BJ) + 4D g \,.
\end{aligned}
\end{equation}
Solving the Hamiltonian (\ref{eq:Hamg}) for $P_g^2$ gives
\begin{equation}
  P_g^2 
     = \frac{Q_3(g)}{4g^2}
\end{equation}
so that the action integral  (\ref{ActGamma}) is obtained from integrating the canonical 
form $P_g \dee g$ over a path with constant $\phi$.

\section{Rotation Number}

We want to check the isoenergetic non-degeneracy condition of the KAM theorem. 
A torus is non-degenerate in this sense if the map from the actions restricted
to a constant energy surface $H=h$ to the frequency ratios $\omega_1 : \omega_2$
is non-degenerate. This means that the frequency ratio (or rotation number) 
$W = \omega_1/\omega_2$
changes when the torus is changed at constant
energy. On a local transversal Poincar\'e section this condition is called 
twist condition.

In our case this is equivalent to the non-vanishing of the partial derivative 
of the rotation number $W$ with respect to the action $J_{1}$. 
By definition the winding number is the ratio of frequences $\omega_{1}, \omega_{2}$, corresponding to the actions $J_{1}$ and $J_2$. If the Hamiltonian is 
expressed in terms of $J=J_1$ and $J_2$ then $\partial_{1} H(J_1, J_2) = \omega_1$ 
and $\partial_{2} H(J_1, J_2) = \omega_2$. Therefore we find by implicit differentiation 
of $J_2(h,j)=j_2$ that
\begin{equation}
W = \frac{\omega_{1}}{\omega_{2}} = -\frac{\partial J_2}{\partial j}.
\end{equation}

However, the simplest way to obtain $W$ is to observe that it is the
advance of the angle $\phi$ conjugate to $J$ during the time of
a full period of the motion of $g = \Gamma_3$. 
The period of the motion is obtained from the reduced equation
of motion $\dot \Gamma_3 = -\Gamma_4$.
On $P_j$ this gives
\begin{equation}
    \dot\Gamma_3^2 = 4 \Gamma_2 \Gamma_3 - j^2
\end{equation}
and eliminating $\Gamma_2$ by using $H_j = h$ gives
\begin{equation}
   \left( \frac{ \dee \Gamma_3}{\dee t}\right)^2 = 4\Gamma_3 \Gamma_2(\Gamma_3;h,j) - j^2\,.
\end{equation}
By separation of variables we obtain the period of the reduced motion as
\begin{equation}
  T(h,j) = \oint \frac{\dee \Gamma_3}{\sqrt{ 4 \Gamma_3 \Gamma_2 - j^2}} = \oint \frac{\dee z}{w} \,.
\end{equation}
To obtain the advance of $\phi$ in time $T$ we change the time $t$ in (\ref{phidot})
to ``time'' $\Gamma_3$ and find
\begin{equation}
\frac{\dee \phi}{\dee \Gamma_3} = \frac{j+2\Gamma_3(\omega +2Cj) + B(2\Gamma_3)^2}{2\Gamma_3 \Gamma_4}
\end{equation}
Expressing $\Gamma_4$ in terms of $\Gamma_3$ on the reduced phase space 
$P_j$ as before the period of the solution of this equation gives the rotation number
$2\pi W$.
The rotation number $W$ can therefore be written as a linear combination of integrals
of the first, second and third kind,
\begin{equation}
2\pi W(h,j) = 
(\omega + 2 C j)   \oint \frac{\dee z }{w} + 
2 B                           \oint \frac{z \, \dee z}{w} + 
\frac{j}{2}               \oint \frac{\dee z}{zw} \,.
\end{equation}
The first integral is of the first kind and proportional to the period $T$.
The last integral is of the third kind and propotional to the action $J_2$.
When $D > 0$ the polynomial $Q_3$ has three real roots, 
which we denote  by $z_\text{neg}, z_\text{min}, z_\text{max}$, 
such that $z_\text{neg} \le 0 \le z_\text{min} \le z_\text{max}$.  
The closed loop integrals encircle the  finite range of positive $Q_3$, and 
therefore can be rewritten by the rule
\begin{equation}
\oint = 2\int\limits_{z_\text{min}}^{z_\text{max}} \,.
\end{equation}  
The elliptic integrals can be transformed to Legendre standard integrals
$K(k)$, $E(k)$, and $\Pi(k)$ of the first, second, and third kind, respectively,
with modulus $k$ and characteristic (or parameter) $n$ given by
\begin{equation} \label{eq:kn}
k^{2} = \frac{z_\text{max} - z_\text{min}}{z_\text{max} - z_\text{neg}},
\quad n = \frac{z_\text{max} - z_\text{min}}{z_\text{max}} \,.
\end{equation}
The result is 
\begin{equation} \label{eq:Wjh}
W(j,h) = \frac{(2B z_\text{neg} + \omega + 2 C j) K(k) + 
2B(z_\text{max} - z_\text{neg}) E(k) + j\Pi(n,k)/({2z_\text{max}})}{\pi \sqrt{2D}\sqrt{z_\text{max} - z_\text{neg}}}.
\end{equation}
Explicit formulas for the vanishing of the twist $\partial W/\partial j$ can be 
derived from this. 

The rotation number is a complicated function of the constants of motion $(h,j)$.
The level lines of this function are shown in Fig.~\ref{fig:spirals} and \ref{fig:spirals2}.
Near the cases where the discriminant of the elliptic curve $E$ defined by 
$Q_3$ in (\ref{Ecurve}) vanishes, simpler formulas can be derived. This 
occurs either at the boundary of the image of the energy-momentum map
described by (\ref{hjofs}), or at the isolated focus-focus point $(h,j) = (0,0)$ inside
the image. In the next section we will treat the latter case.

\section{Rotation Number near the Focus-Focus Point}

\begin{figure}
\centerline{\includegraphics[width=10cm, height=6cm]{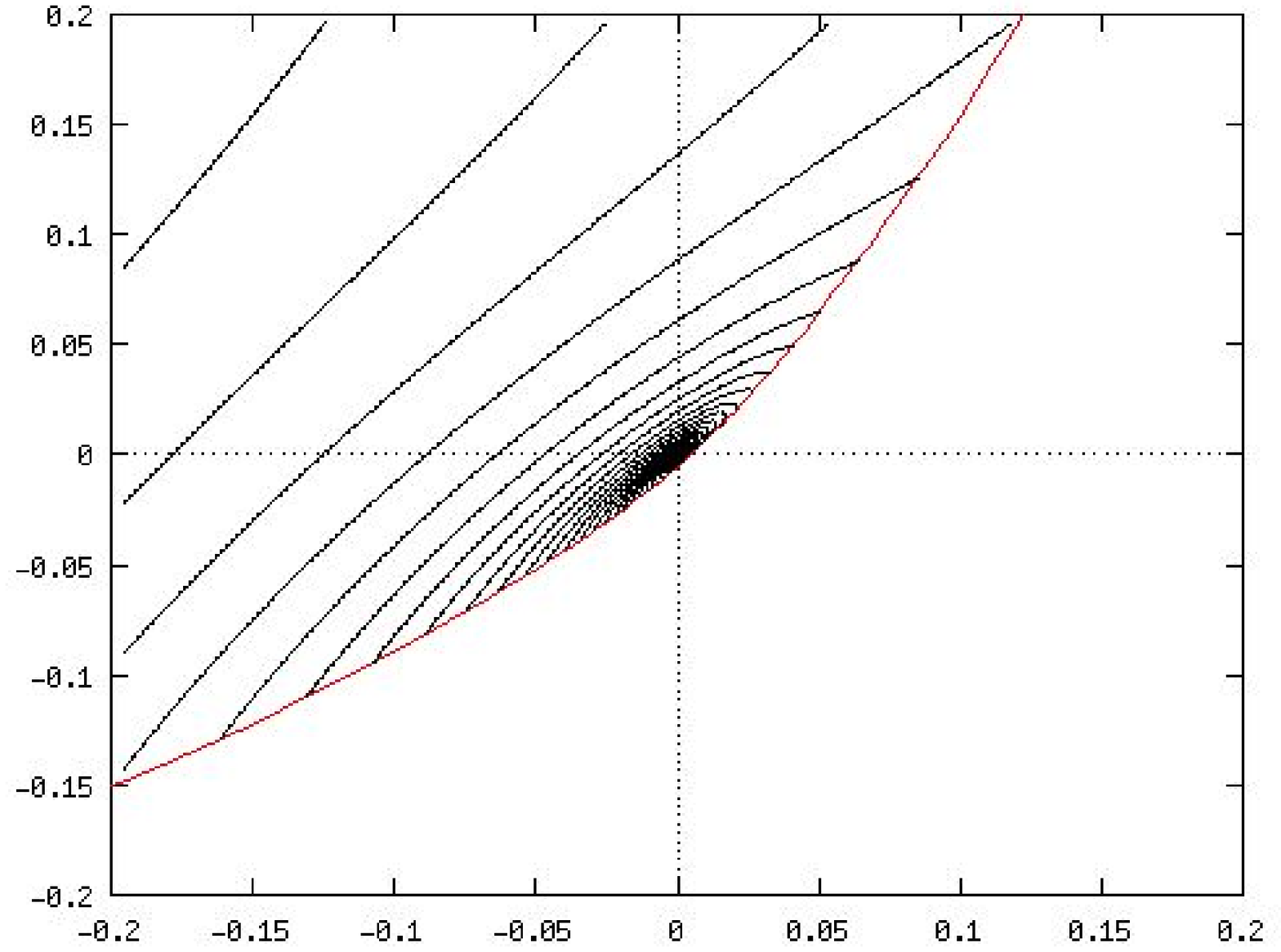}}
\centerline{\includegraphics[width=10cm, height=6cm]{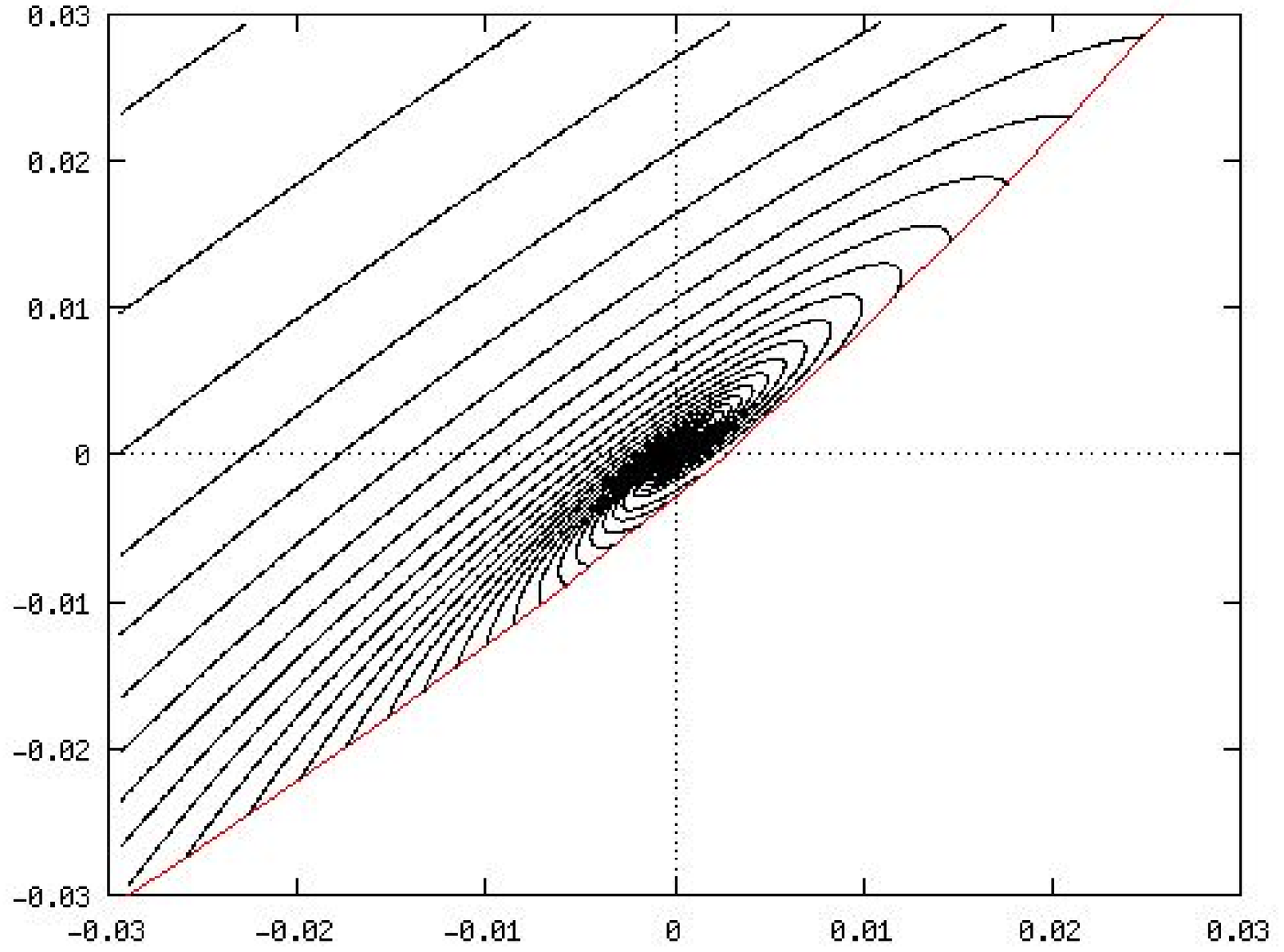}}
\centerline{\includegraphics[width=10cm, height=6cm]{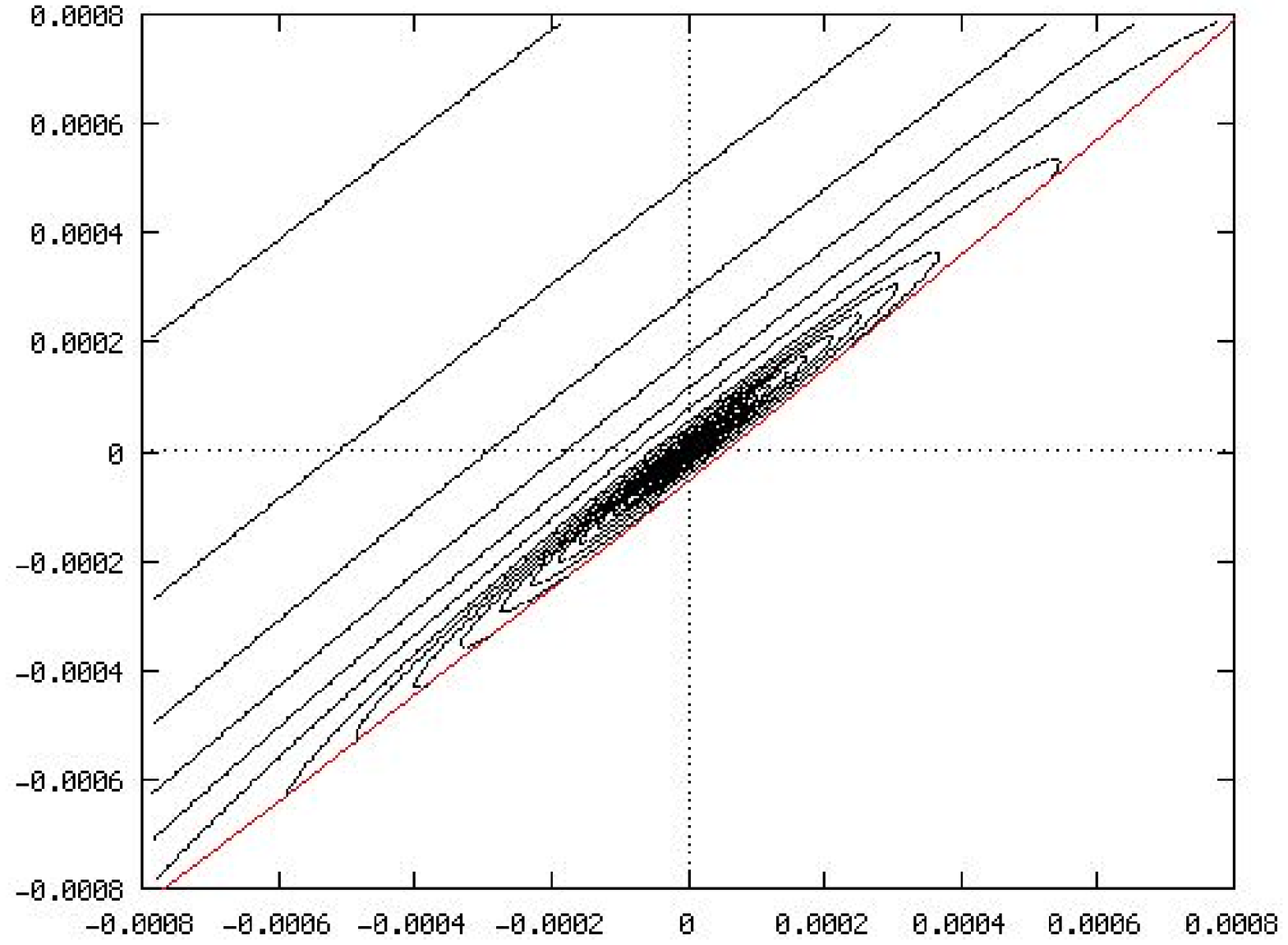}}
\caption{The level lines of the rotation number on the $(j,h)$ plane.
The bifurcation parameter $\delta$ is decreasing from top to bottom, 
$\delta = -0.2, -0.15,  -0.02$ ($\omega = B = C = -D = 1$). The level lines are spirals. 
The boundary of the image of the energy momentum map given by 
a family of relative equilibria is also shown.} \label{fig:spirals}
\end{figure}
\begin{figure}
\centerline{\includegraphics[width=10cm, height=6cm]{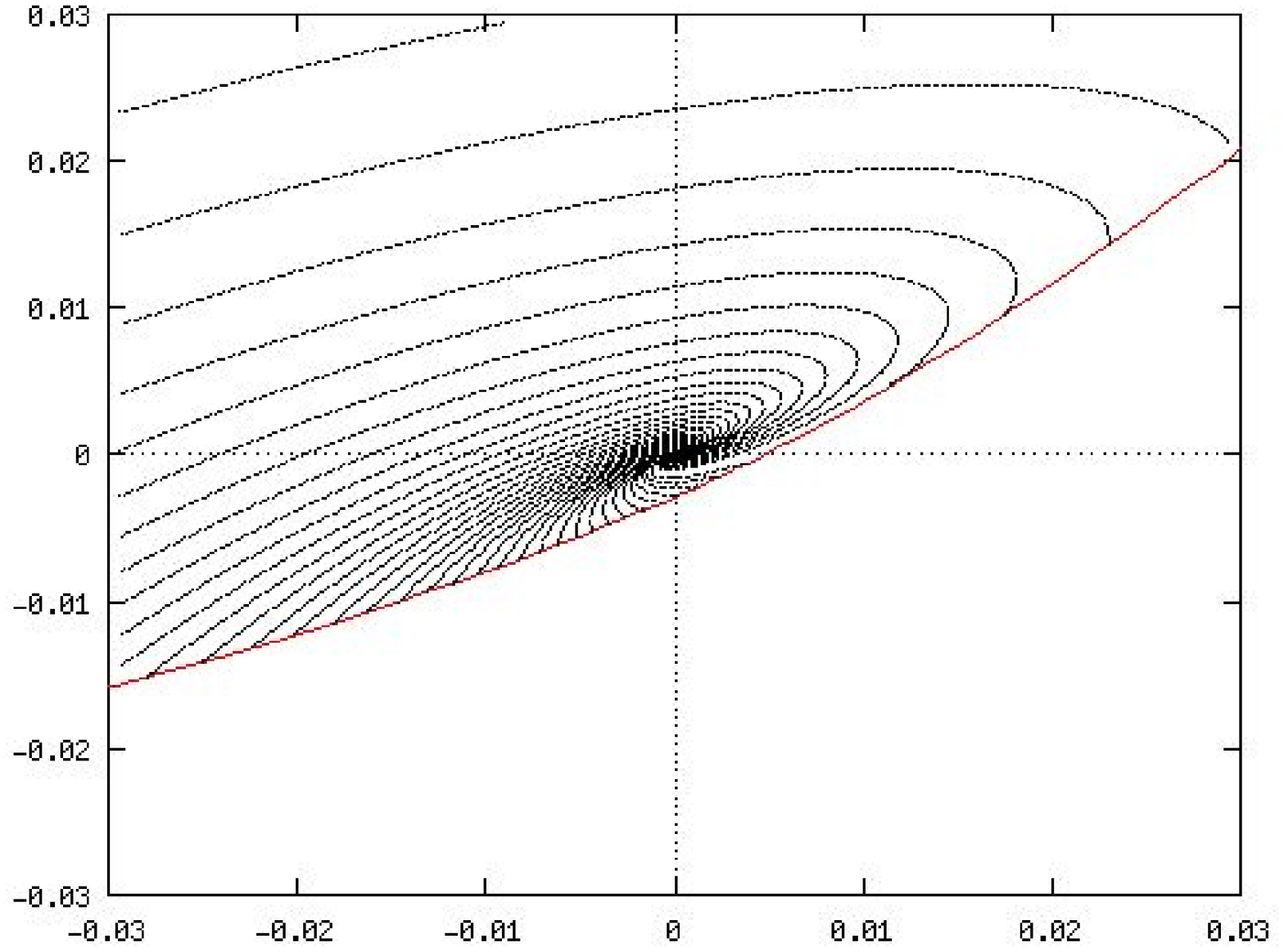}}
\centerline{\includegraphics[width=10cm, height=6cm]{fig_m015M2}}
\centerline{\includegraphics[width=10cm, height=6cm]{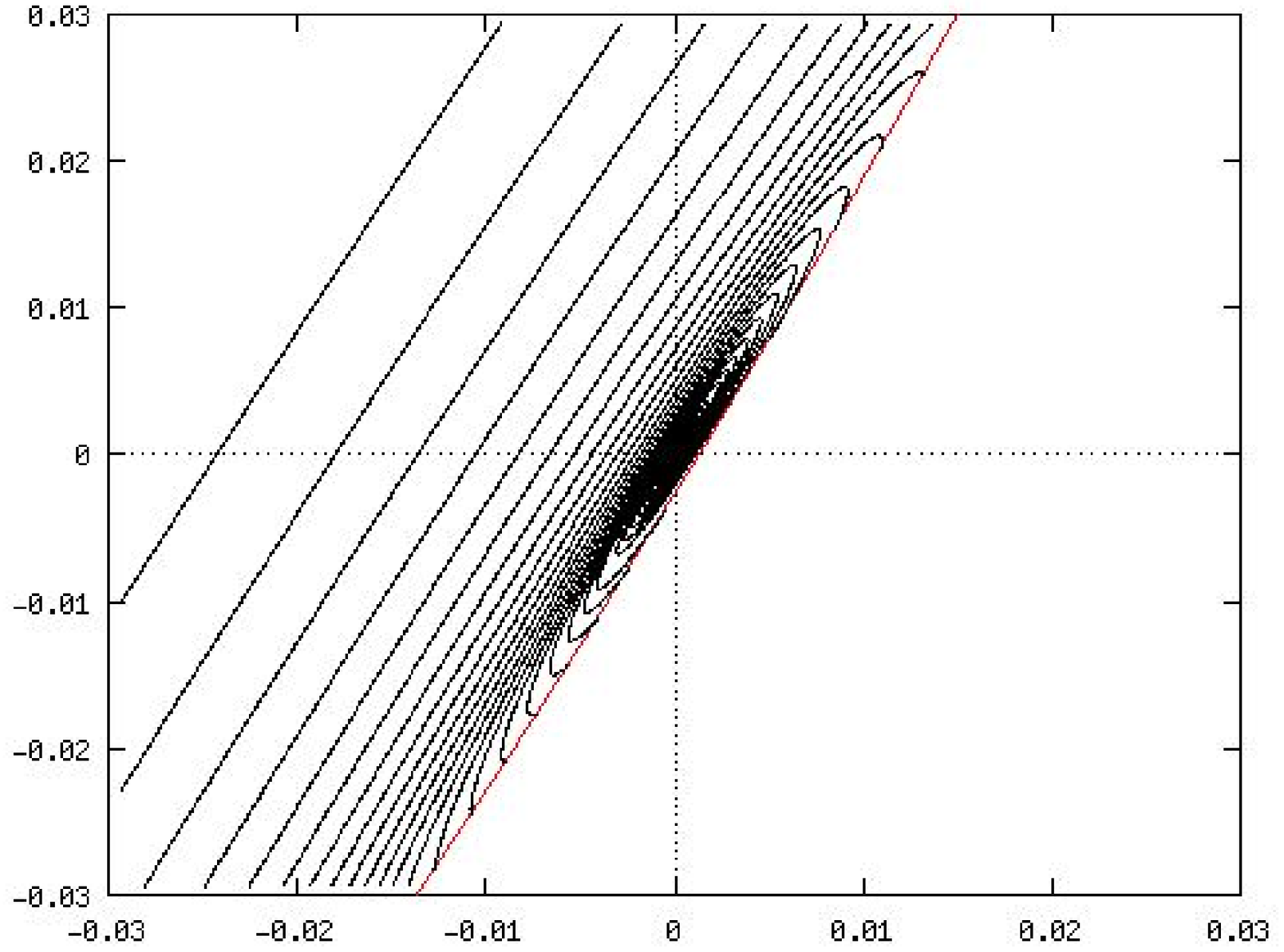}}
\caption{The level lines of the rotation number on the $(j,h)$ plane.
Here the imaginary part of the eigenvalue $\omega$ is changed, from 
top to bottom $\omega = 1/2, 1, 2$ ($\delta = -0.15, B = C = -D =1$).} \label{fig:spirals2}
\end{figure}

We introduce a small parameter by scaling $h$ and $j$ by epsilon, 
hence replace $h \to h\eps$, $j \to j \eps$. This means that we obtain 
an expansion that approaches the origin on a ray. Alternatively one can view
$\eps$ as a formal expansion parameter that keeps track of the fact that both,
$h$ and $j$ are small and of the same order. The focus point only exists for
$\delta < 0$, which we henceforth assume.
At the origin $Q_3$ reduces to 
\begin{equation}
Q_3(z; h=0, j=0) = -4 z^2 ( 2 D z + \delta)\,,
\end{equation}
so that $z_\text{min}$ and $z_\text{neg}$ collide at 0 and 
$z_\text{max} = -\delta/2/D$. The roots can be expanded in power series
in  $\eps$, and the result is
\begin{equation}
\begin{aligned}
2z_\text{neg}  & = - \frac{l+\rho}{\alpha}\eps + O(\eps^2), \\
2z_\text{min} & = - \frac{l-\rho}{\alpha} \eps + O(\eps^2), \\
2z_\text{max} & = \frac{\alpha^2}{D} + 2 \frac{Dl-jB\alpha}{D\alpha}\eps + O(\eps^2) \,.
\end{aligned}
\end{equation}
Here and in the following we use the abbreviations
\begin{equation}
    l = \frac{h - \omega j}{\alpha}, \quad
    \rho^2 = l^2 +  j^2, \quad
    f = \frac{D}{\alpha^3}, \quad 
    \alpha = \sqrt{-\delta} \,.
 \end{equation}
The expressions are only real in the case $\delta < 0$, otherwise the focus-focus
point does not exist.
Inserting this into (\ref{eq:kn}) gives
\begin{equation} \label{eq:knser}
\begin{aligned}
k^{2} & = 1 - 2 f \rho \eps + O(\eps^2), \\
       n & = 1 +f (l  -  \rho)\eps + O(\eps^2).
\end{aligned}
\end{equation}
For small $\eps$ both, $k$ and $n$ are close to 1 and they satisfy the inequality
\begin{equation} \label{eq:k2len}
  k^2 \le n \le 1 \,.
\end{equation}
In the limit $k \to 1$ the elliptic integrals are singular, but there are expansions
that include the logarithmically diverging terms. The details of this expansion
can be found in the appendix. The result for the rotation number is
\begin{equation} \label{eq:Wexpansion}
   2\pi W(h,j) = -\frac{\omega}{ \alpha } \ln \rho - \tan^{-1}\frac{j}{l} + O(1)\,.
\end{equation}
Keeping terms only up to order 1 is enough because when the twist condition
$\partial W / \partial j$ is calculated the present terms both give singular
contributions, the constant term disappears and the first order term in $\eps$
is very small compared to the singular terms.
Note that $W$ is not a single valued function. The fact that $\tan^{-1}(j/l)/2\pi$ 
changes by one when the origin is encircled is an expression of the monodromy
of this focus-focus point, see \cite{Duistermaat98}. The expression given by 
the elliptic integrals, see the appendix, gives a continuous function 
which is, however, not differentiable when $j=0$.

The level lines of $W$ are spirals, which are easily parametrized 
in polar coordinates with the radius $\rho$ as parameter.
Instead of viewing these spirals as the level lines of $W$ where 
$W$ is a many-valued function, it might be easier to see them 
as the integral curves of a flow in the plane that has an equilibrium
point at $h=j=0$. The linear approximation to this flow is best
written in complex notation $z = l + \imag j$ so that $\dot z = \lambda z$
where $\lambda = \alpha + \imag \omega$. Therefore we may 
formulate the result like this: {\em The level 
lines of $W$ are the integral curves of a planar node with the same
eigenvalue as the focus-focus point.}

Calculating the leading order condition for the vanishing of twist is easily 
done by differentiating the first two terms of the expansion with respect to $j$.
The result is that 
\begin{equation}
  \frac{\partial W}{\partial j} = 0 \quad  \Rightarrow    \quad 
     h (\omega^2 - \alpha^2) = j \omega (\omega^2 + \alpha^2) \,.
\end{equation}
This means that the twist vanishes on a line that has a tangent whose slope
at the focus point can be read of from the above expression as
\begin{equation}
   \frac{h}{j} = \omega\frac{\omega^2 + \alpha^2}{\omega^2 - \alpha^2} \,.
\end{equation}
In particular this slope only depends on the eigenvalue of the focus-focus point.
Approaching the bifurcation point $\alpha \to 0$, and hence the slope 
approaches $\omega$, see Fig.~\ref{fig:spirals} and \ref{fig:spirals2}.

\section{Vanishing Twist of Periodic Orbits}

Now we consider the vanishing of the twist for the relative equilibria
$(h_\text{crit}(s), j_\text{crit}(s))$ in the boundary of the image of the energy-momentum 
map. 
First we rescale the parameters, in order to keep the formulas manageable:
\begin{equation}
B = \frac{D b}{\omega},\ \ C = \frac{D c}{\omega^{2}},\ \  
\mu = \frac{\delta}{\omega^2}, \ \ 
j = \frac{\hat j \omega^{3}}{D},\ \ h = \frac{\hat h \omega^{4}}{D},\ \  z = \frac{\zeta \omega^{2}}{D}.
\end{equation}
Note that when $\delta < 0$, which is the case we are interested here, 
we have $\mu = -\alpha^2/\omega^2$.
In terms of the new parameters the polynomial $Q(z)$ becomes
\begin{equation}
Q(z) = \frac{\omega^{6}}{D^{3}}\hat Q(\zeta),\ \  
\hat Q(\zeta) = - (\zeta^{3} + (\mu + 2 b \hat j)\zeta^{2} + 2(\hat j + c \hat j^{2} - \hat h)\zeta + \hat j^{2}),
\end{equation}
and the derivative $\partial W/\partial j$ is
\begin{equation} \label{eq:dWdjscaled}
\frac{\partial W}{\partial j}(j,h) = \frac{D}{4\pi\omega^{3}}\left(
\oint \frac{(1 + 2 c \zeta){\rm d}\zeta}{\zeta \hat Q^{1/2}(\zeta)} + \oint \frac{(b \zeta^{2} + (1 + 2 c \hat j)\zeta + \hat j)^{2}{\rm d}\zeta}{\zeta \hat Q^{3/2}(\zeta)}\right).
\end{equation}
The parametrisation of the critical values has already been obtained in (\ref{hjofs}).
After the scaling it reads
\begin{equation} \label{hjofsscaled}
\hat d(s) = \frac{\mu - s^2}{2(bs + 1)}, \quad
\hat j_\text{crit}(s) = s \hat d(s), \quad
\hat h_\text{crit} = \omega \hat j_\text{crit}(s) + c\hat j^2_\text{crit}(s) + 2\hat d(s)^2 + 2s \hat j_\text{crit}(s) \,. 
\end{equation}
On the curve $(\hat j_\text{crit}(s), \hat h_\text{crit}(s))$ the
 integrals on the right hand side of (\ref{eq:dWdjscaled}) 
can be computed by the method of residues. Replacing 
$\hat j$ and $\hat h$ wherever they appear by their critical values parametrised by 
$s$ gives a condition for the  vanishing of the twist on the curve of critical values. 
It implies the following polynomial equation
\begin{equation} \label{eq:R9spo}
R_{9}(s) = \sum\limits_{k=0}^{9} a_{k} s^{k} = 0,
\end{equation}
where
\begin{equation}
\begin{aligned}
a_{0} =& -12\mu - 4(4+b)\mu^2 - ( 3b^{2} - 16c)\mu^{3}, \\
a_{1} =& -24(5+b)\mu - 4(b + b^{2} - 6c)\mu^{2} + 4bc\mu^{3},  \\
a_{2} =& -84 - 12(1 + 14 b + b^{2})\mu  + 3(15b^{2} - 8c + 8bc)\mu^{2} - 12c^{2}\mu^{2},  \\
a_{3} =& 4(30 - 66b) + 16(b + b^{2} + 9c)\mu + (24 b^{3} - 52bc)\mu^{2},  \\
a_{4} =& 12(5 + 25b - 23b^{2}) + (64b^{3} - 29b^{2} + 192c + 336bc)\mu - 4c(16b^{2} + 15c)\mu^{2},  \\
a_{5} =& -12(8b^{3} - 31b^{2} - 7b + 14c) - 12b(12b^{2} - 29c - 16bc)\mu - 96bc^{2}\mu^{2},  \\
a_{6} =& 320b^{3} + 83b^{2} - 312c - 360bc - (96b^{4} - 64b^{2}c - 156c^{2})\mu,  \\
a_{7} =& 4b(32b^{3} + 38b^{2} - 171c - 48bc) - 64bc(2b^{2} - 3c)\mu,  \\
a_{8} =& 12(8b^{4} - 32b^{2}c - 7c^{2}),  \\
a_{9} =& -96bc^{2}.
\end{aligned}
\end{equation}
The first two coefficients $a_{0}$ and $a_{1}$ are of the first order in $\mu$, while 
$a_{2}$ is non-zero for small $\mu$. Therefore the roots of $R_9$ can be 
expanded in powers of $|\mu|^{1/2} = \alpha/\omega$, 
and for small $|\mu|$ there are only
two roots in a $|\mu|^{1/2}$ neighborhood of the origin.
The result is
\begin{equation}
s_{\pm} = 
         \pm\sqrt{\frac{-\mu}{7}} + \frac47(10-b)\mu  + O(\vert \mu \vert^{3/2}).
\end{equation}
%
The critical values of the twistless periodic orbit are then given by
\begin{align}
\hat j_{\pm} &= \pm\frac47 \sqrt{ \frac{-\mu^3}{7}} + \frac{8}{7^3}(25-6b)\mu^2 + 
               O(\vert\delta\vert^{5/2}),  \\
\hat h_{\pm} & = \pm\frac47 \sqrt{ \frac{-\mu^3}{7}} + \frac{8}{7^3}(43/2-6b)\mu^2 + 
               O(\vert\delta\vert^{5/2}) \,.
\end{align}
This gives two points in each of the Figures~\ref{fig:spirals} and \ref{fig:spirals2}, 
at which the curve of vanishing twist emanating from the origin crosses the boundary
of the image of the energy-momentum map. The value of the rotation number 
at these points is
\begin{equation}
 \sqrt{5} W_\pm = \frac12 \sqrt{ \frac{7}{-\mu} } \pm \frac17 (5 - 2b) + O(|\mu|^{1/2})\,.
\end{equation}

We have now treated both limiting cases, that near the focus-focus 
point, and that near the elliptic relative equilibria. 
The curve of vanishing twist for all values in between can in principle 
be computed from the derivative of  (\ref{eq:Wjh}), 
and will connect the results from the two limiting cases for small 
values of $\delta$.

\appendix

\section{Expansion of $W$}

We need to expand the Legendre standard integrals $K$, $E$, and $\Pi$ 
in the limit $k \to 1$. For the expansion of the integral of the third kind it is important to 
take the inequality $k^2 \le n \le 1$  (\ref{eq:k2len}) into account.
In this limit the following formulas can e.g.\ be found in \cite{AS}:
\begin{equation} \label{eqn:Piex}
          \Pi(n,k) = K(k) + \frac{1}{2} \pi {\cal R} (1 - \Lambda_{0}(\theta,k)),
\end{equation}
\begin{equation}
{\cal R} = \left(\frac{n}{(1 - n)(n - k^2)}\right)^{1/2}
              = \left( \frac{(z_\text{neg}-z_\text{max})z_\text{max}}{z_\text{neg} z_\text{min}} \right)^{1/2} \,.
\end{equation}
Here $\Lambda_0$ is Heumann's Lambda function, which can be expressed 
in terms of incomplete elliptic integrals $F(\theta,k)$ and $E(\theta,k)$
of the first and the second kind, respectively, as
\begin{equation}
      \Lambda_{0}(\theta,k) = \frac{2}{\pi}\left(K(k)E(\theta,k') - (K(k) - E(k))F(\theta,k')\right),
\end{equation}
\begin{equation}
       \theta = \arcsin\sqrt{\frac{1 - n}{1 - k^{2}}}
                  = \arcsin\sqrt{\frac{z_\text{min}(z_\text{max} - z_\text{neg})}{z_\text{max}(z_\text{min}-z_\text{neg})}} \,.
\end{equation}
Note that the complementary modulus $k' = \sqrt{1-k^2}$ and the parameter satisfy 
\[
   k'^2 = 1 - k^2  = O(\eps), \quad 1-n = O(\eps) \,,
\]
see (\ref{eq:knser}). Accordingly $k'^2$ and $1-n$ are of order $\eps$, 
while $\theta$ is not small but of order 1. 
The prefactor $\cal R$ in (\ref{eqn:Piex}) cancels with the prefactor of $\Pi$ in (\ref{eq:Wjh}), 
up to a factor of $1/2$. Therefore we find the (still exact) formula
\begin{equation}
 W = c_1 K + c_2 E + \frac12 ( 1 - \Lambda_0(\theta,k) )
\end{equation}
where with $\sqrt{ z_\text{max} - z_\text{neg} } =  \alpha/\sqrt{2D} + O(\eps)$ it follows
\begin{align}
    \pi c_1 &= \frac{\omega + 2B z_\text{neg} + 2Cj + j/(2z_\text{max})}{ \sqrt{2D} \sqrt{z_\text{max} - z_\text{neg}}}
               =  \frac{\omega}{\alpha} + O(\eps), \\
    \pi c_2 &= \frac{2B\sqrt{z_\text{max} - z_\text{neg}}}{\sqrt{2D}}
               = \frac{B \alpha }{ D} + O(\eps) \,.
\end{align}
The elliptic integrals in the limit $k' \to 0$ have a logarithmic divergence
of leading order
\begin{equation}
          \Lambda = \log\frac{4}{k'}\,.
\end{equation}
The convergent expansions in this limit are
\begin{equation}
          K(k) = \Lambda + \frac{\Lambda - 1}{4} k'^{2} + O(\Lambda k'^{4}),
\end{equation}
\begin{equation}
        E(k) = 1 + \frac{1}{2}\left(\Lambda - \frac{1}{2}\right) k'^{2} + O(\Lambda k'^{4}),
\end{equation}
The incomplete elliptic integrals of modulus $k'$ have regular expansions
since $k' \to 0$, so that
\begin{equation}
F(\theta,k') - E(\theta,k') = \int\limits_{0}^{\theta}\frac{k'^{2}\sin^{2}\varphi{\rm d}\varphi}{\sqrt{1 - k'^{2}\sin^{2}\varphi}} = \left(\frac{\theta}{2} - \frac{\sin2\theta}{4}\right)k'^{2} + O(k'^{4}),
\end{equation}
\begin{equation}
F(\theta,k') = \int\limits_{0}^{\theta}\frac{{\rm d}\varphi}{\sqrt{1 - k'^{2}\sin^{2}\varphi}} = \theta + \frac{1}{2}\left(\frac{\theta}{2} - \frac{\sin2\theta}{4}\right)k'^{2} + O(k'^{4})\,.
\end{equation}
For the Heumann Lambda function this gives
\begin{equation}
\Lambda_{0}(\theta,k) = \frac{2}{\pi}\left(\theta + \frac{1}{4}\left(\Lambda - \frac{1}{2}\right)\sin(2\theta) k'^{2} + O(\Lambda k'^{4})\right).
\end{equation}
The leading order terms in the expansion of (\ref{eq:Wjh}) 
come from the diverging $K$. Since $E \to 1$ for 
$k \to 1$ there is only a constant contribution from $c_2$. From $\Lambda_0$ the
leading term is merely $\theta$, so that all together
\begin{equation}
   \pi W = \frac{\omega}{\alpha} \Lambda + \frac{\pi}{2}  - \theta  + \frac{B\alpha}{D} + O(\eps)
\end{equation}
It remains to understand the parameter dependence of $\theta$. 
Expanding $\theta$ gives
\begin{equation}
   2 \theta  \approx 2 \arcsin\sqrt{\frac{\rho-l}{2\rho}} + O(\eps).
\end{equation}
This can be simplified using the relation
\begin{equation}
   2 \arcsin {\beta} = \arctan {\gamma} \quad \Rightarrow \quad
    \gamma = 2 \frac{\sqrt{1-\beta^2}\beta}{1-2\beta^2} \,.
\end{equation}
Inserting $\beta^2 = (\rho - l)/(2\rho)$ and using $\rho^2 = l^2 + j^2$ gives
$\gamma = j/l$, so that (\ref{eq:Wexpansion}) is proved.
Note that when $j=0$ we have $\rho = l$ and the root $z_\text{min}$ 
collides with the pole at $z = 0$ in the third kind integral. 
This is the place where the dependence on the parameters of $\Pi(n,k)$
is continuous, but not smooth. 


\bibliographystyle{plain}
\bibliography{../notwist,../../bib_cv/all}

\def\cprime{$'$}
\begin{thebibliography}{10}

\bibitem{AS}
Milton Abramowitz and Irene~A. Stegun, editors.
\newblock {\em Handbook of mathematical functions with formulas, graphs, and
  mathematical tables}.
\newblock Dover Publications Inc., New York, 1992.
\newblock Reprint of the 1972 edition.

\bibitem{AKN}
V.~I. Arnold, V.~V. Kozlov, and A.~I. Neishtadt.
\newblock {\em Mathematical aspects of classical and celestial mechanics}.
\newblock Springer-Verlag, Berlin, 1997.
\newblock Translated from the 1985 Russian original by A. Iacob, Reprint of the
  original English edition from the series Encyclopaedia of Mathematical
  Sciences [{\it Dynamical systems. III}, Encyclopaedia Math. Sci., 3,
  Springer, Berlin, 1993; MR 95d:58043a].

\bibitem{CushmanBates97}
Richard~H. Cushman and Larry~M. Bates.
\newblock {\em Global aspects of classical integrable systems}.
\newblock Birkh\"auser Verlag, Basel, 1997.

\bibitem{CGM96}
D.~del Castillo-Negrette, J.M. Greene, and P.J. Morrison.
\newblock Area preserving nontwist maps: Periodic orbits and transition to
  chaos.
\newblock {\em Physica D}, 91(1):1--23, 1996.

\bibitem{DL98}
A.~Delshams and R.~de~la Llave.
\newblock K{AM} theory and a partial justification of {G}reene's criterion for
  nontwist maps.
\newblock {\em SIAM J. Math. Anal.}, 31(6):1235--1269 (electronic), 2000.

\bibitem{Duistermaat84}
J.~J. Duistermaat.
\newblock Bifurcation of periodic solutions near equilibrium points of
  {H}amiltonian systems.
\newblock In {\em Bifurcation theory and applications (Montecatini, 1983)},
  volume 1057 of {\em Lecture Notes in Math.}, pages 57--105. Springer, Berlin,
  1984.

\bibitem{Duistermaat98}
J.~J. Duistermaat.
\newblock The monodromy in the {H}amiltonian {H}opf bifurcation.
\newblock {\em Z. Angew. Math. Phys.}, 49(1):156--161, 1998.

\bibitem{DI03a}
H.~R. Dullin and A.~V. Ivanov.
\newblock (Vanishing) twist in the saddle-centre and period-doubling
  bifurcation.
\newblock {\em (preprint http://arxiv.org/abs/nlin.CD/0305033)}, 2003.

\bibitem{DM02}
H.~R. Dullin and J.~D. Meiss.
\newblock Twist singularities for symplectic maps.
\newblock {\em Chaos}, 13(1):1--16, 2003.

\bibitem{DMS98b}
H.~R. Dullin, J.~D. Meiss, and D.~G. Sterling.
\newblock Generic twistless bifurcations.
\newblock {\em Nonlinearity}, 13:203--224, 2000.

\bibitem{HowHum95}
J.~E. Howard and J.~Humpherys.
\newblock Nonmonotonic twist maps.
\newblock {\em Physica D}, 80(3):256--276, 1995.

\bibitem{HowHoh84}
J.E. Howard and S.M. Hohs.
\newblock Stochasticity and reconnection in hamiltonian systems.
\newblock {\em Physical Review A}, 29:418, 1984.

\bibitem{Moeckel90}
R.~Moeckel.
\newblock Generic bifurcations of the twist coefficient.
\newblock {\em Ergodic Theory Dynam. Systems}, 10(1):185--195, 1990.

\bibitem{MRS88}
J.~A. Montaldi, R.~M. Roberts, and I.~N. Stewart.
\newblock Periodic solutions near equilibria of symmetric {H}amiltonian
  systems.
\newblock {\em Philos. Trans. Roy. Soc. London Ser. A}, 325(1584):237--293,
  1988.

\bibitem{Moser76}
J.~Moser.
\newblock Periodic orbits near an equilibrium and a theorem by {A}lan
  {W}einstein.
\newblock {\em Comm. Pure Appl. Math.}, 29(6):724--747, 1976.

\bibitem{Ruessmann87}
Helmut R{\"u}ssmann.
\newblock Nondegeneracy in the perturbation theory of integrable dynamical
  systems.
\newblock In {\em Number theory and dynamical systems (York, 1987)}, volume 134
  of {\em London Math. Soc. Lecture Note Ser.}, pages 5--18. Cambridge Univ.
  Press, Cambridge, 1989.

\bibitem{Simo98}
C.~Sim{\'o}.
\newblock Invariant curves of analytic perturbed nontwist area preserving maps.
\newblock {\em Regular {\&} Chaotic Dynamics}, 3:180--195, 1998.

\bibitem{Sokolskii77}
A.~G. Sokol{\cprime}ski\u{\i}.
\newblock On stability of an autonomous {H}amiltonian system with two degrees
  of freedom under first-order resonance.
\newblock {\em Prikl. Mat. Meh.}, 41(1):24--33, 1977.

\bibitem{vanderMeer85}
Jan-Cees van~der Meer.
\newblock {\em The {H}amiltonian {H}opf bifurcation}, volume 1160 of {\em
  Lecture Notes in Mathematics}.
\newblock Springer-Verlag, Berlin, 1985.

\bibitem{Weinstein73}
Alan Weinstein.
\newblock Normal modes for nonlinear {H}amiltonian systems.
\newblock {\em Invent. Math.}, 20:47--57, 1973.

\end{thebibliography}

\end{document}